# Exploration of Bi-Level PageRank Algorithm for Power Flow Analysis Using Graph Database

Chen Yuan[a], Yi Lu[b], Kewen Liu[c], Guangyi Liu[a], Renchang Dai[a], Zhiwei Wang[a]
[a] Global Energy Interconnection Research Institute North America, San Jose, CA, USA
[b] State Grid Sichuan Electric Power Company, Chengdu, Sichuan, China
[c] Global Energy Interconnection Research Institute, Beijing, China
{chen.yuan@geirina.net, guangyi.liu@geirina.net}

*Abstract*—**Compared with traditional relational database, graph database (GDB) is a natural expression of most real-world systems. Each node in the GDB is not only a storage unit, but also a logic operation unit to implement local computation in parallel. This paper firstly explores the feasibility of power system modeling using GDB. Then a brief introduction of the PageRank algorithm and the feasibility analysis of its application in GDB are presented. Then the proposed GDB based bi-level PageRank algorithm is developed from PageRank algorithm and Gauss-Seidel methodology realize high performance parallel computation. MP 10790 case, and its extensions, MP 10790*10 and MP 10790*100, are tested to verify the proposed method and investigate its parallelism in GDB. Besides, a provincial system, FJ case which include 1425 buses and 1922 branches, is also included in the case study to further prove the proposed algorithm's effectiveness in real world.**

## I. Introduction

Power flow acts as the basic but critical role in the power system analysis. Most of applications in Energy Management Systems (EMS), like state estimation, "N-1" contingency analysis, security constrained economic dispatch, and transient stability, are based on power flow solution. With the large integrations of smart grid technologies [1], [2], renewable energy [3]–[5], and electric vehicles [6], the complexity of the power grid has been drastically increased and more frequent fluctuations are brought into power systems. The traditional EMS cannot well follow the dynamics in modern power grids. A fast computing algorithm for efficiently solving power flow has a profound influence on EMS. Parallel computing is one of promising methods to improve computation efficiency. However, the state of art of power flow analysis does not effectively make use of the parallel capability, since the relational database and computation algorithm used for existing power flow analysis were not specifically designed for parallel computing. With the fast development of computing technology and graph theory based applications, graph based high performance computation, *graph computing*, is a feasible option for high-performance parallel computing [7], since it was developed to deal with distributed storage and parallel computing in big data analysis, and applicable to solve complicated scenarios with iterations [8].

Most commercial power system analysis tools use the fast-decoupled method and Newton-Raphson method to solve power flow problem. The fundamental algorithm is based on matrix manipulations and the essence is using two-dimensional sparse matrix to represent system topology. Algorithms for transmission systems power flow, including technologies of sparse matrix [9], sparse vector [10] and node ordering [11], have been well studied. With the evolution of software systems and hardware configurations in parallel computing, the external conditions of the power flow analysis in large-scale systems become mature. Reference [12] used distributed computation technology to implement parallel computation of power flow. Besides, GPU based parallel computing was introduced and applied to power flow calculation [13], [14]. On the other hand, our previous works have investigated the feasibility and the high performance of graph database in power system energy management systems (EMS) applications, like system modeling, state estimation, "N-1" contingency analysis, and security constrained economic dispatch [15]–[19].

In this paper, a bi-level PageRank algorithm using graph database (GDB) is proposed for power flow analysis with parallel computing. Graph is the natural expression of real-world systems. Each node in the GDB is not only a storage unit, but also a logic operation unit to implement local computation in parallel. Power system network modeling on the basis of GDB is firstly investigated. Then the graph based parallel computing application in power flow analysis is carried out by borrowing ideas from PageRank algorithm and Gauss-Seidel method, since both of them are designed to solve non-linear problems and applicable with parallel computing. The proposed strategy ensures high performance computation with the help of node-based parallel computing in GDB, and meanwhile improves the convergence by making use of damping factor and bi-level parallel computing.

This paper is organized as follows: graph database and graph computing will be briefly introduced in Section II, including its applications in power systems. Then the proposed bi-level PageRank algorithm for power flow analysis using graph database is well elaborated in Section III. Section IV verifies the proposed algorithm accuracy using IEEE standard cases and presents its high performance in multiple threads using large real systems. At last, the paper is concluded in Section V and future work is also discussed in this section.

This work was supported by State Grid Corporation technology project SGRIJSKJ(2016)800.

## II. Graph Database and Graph Computing

### *A. Graph and Graph Database*

Graph is a data structure modeling pairwise relations between objects in a network. Many real-life scenarios can be modeled as graphs. In mathematics, a graph is represented as $G = (V, E)$, in which $V$ indicates a set of vertices, representing objects, within the graph, $G$, and the set of edges, standing for relations between objects, in the graph is represented as $E$, expressing how these vertices relate to each other. Each edge is denoted by $e = (i, j)$ in $E$, where we refer to $i$ in $V$ and $j$ in $V$ as head and tail of the edge $e$, respectively. Graph theory is the study of graphs, which are used to model relations between objects. A graph could be either directed or undirected. For an undirected graph, the relations are bi-directional between each two connected vertices and there is no distinction in both directions. While a directed graph has directed edges starting from one vertex to another.

GDB uses graph structures for semantic queries with vertices, edges and attributes to represent and store data in vertices and edges. Such database allows data in the store to be linked together directly and retrieved with one graph operation. So, compared with relational database (RDB), which is based on the relational model to store data, GDB permits managing data in its natural structure. Fig. 1 provides a comparison between RDB and GDB, taking an example of IEEE 5-bus system. In RDB, there exists redundant storage in bridge tables, providing join functions by using common attributes. In addition, the data search is complicated through join operations and the time consumption exponentially increases with the database size. However, GDB is very different. It is a natural expression of a real-world system. No join operation is needed, and data are directly stored as attributes in nodes and edges. For example, in Fig. 1, the 5-bus system keeps the same topology in its GDB and system information are respectively distributed to vertices' attributes and edges' attributes. Then, the operations related to data search are more convenient with graph traversal. A testing on an open-source GDB management system, Neo4j, against a widely used RDB management system, MySQL, shows that the overall performance of data search in Neo4j is much better than MySQL [20].

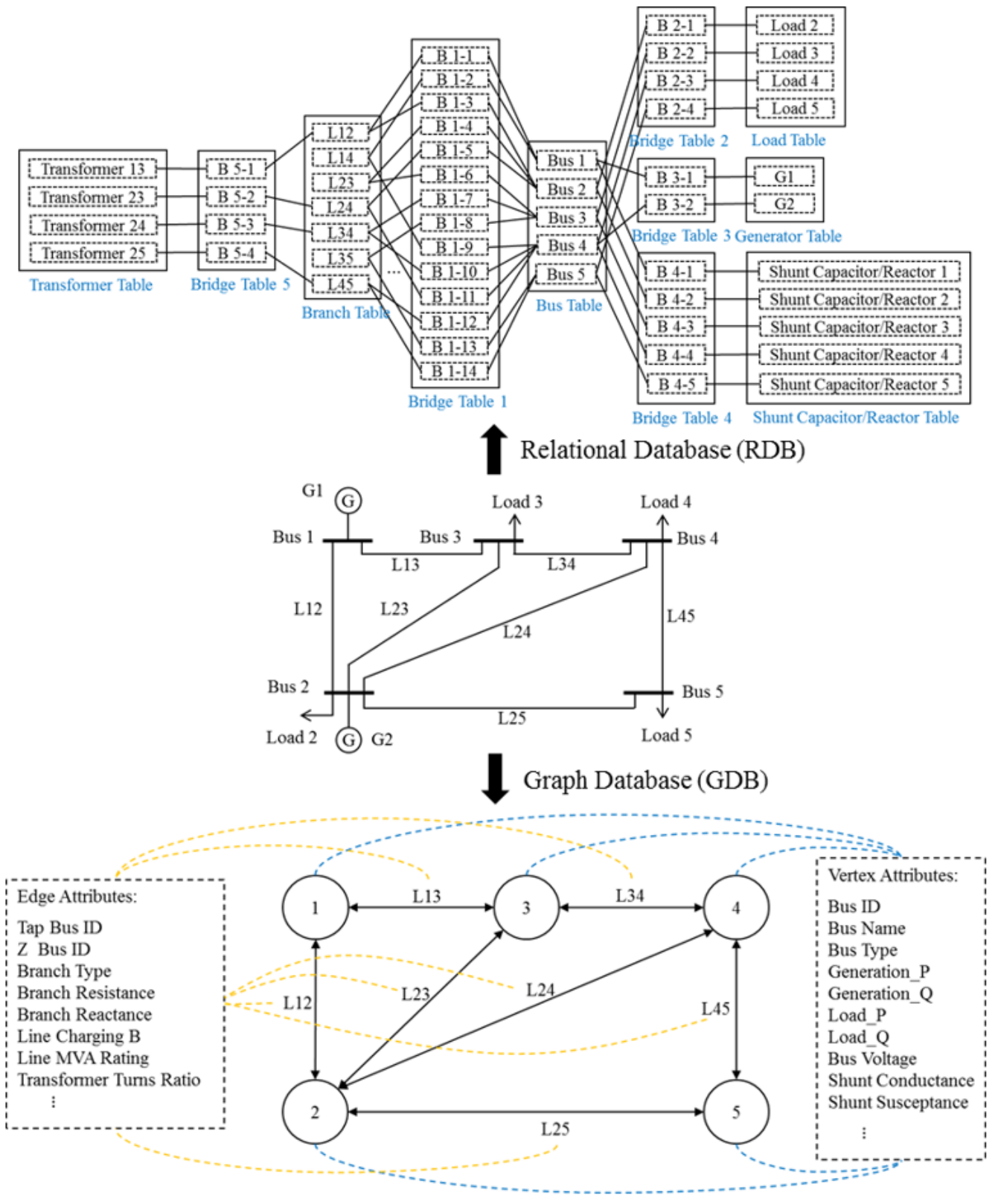

Figure 1. A comparison between RDB and GDB in power systems modeling

### *B. Graph Database Applications in Power Systems*

According to the GDB description and the comparison between GDB and RDB in Section II.A, this paper investigates the GDB applications in power systems. The mapping between graph and power system is presented in Fig. 2. A 6-bus power system network is converted to a graph containing 6 vertices and 7 undirected edges. Both have the same structure. In a n-bus power system, its admittance matrix is a $n \times n$ symmetrical matrix. It not only represents the nodal admittance of the buses, but also displays the topology structure of this power system. That is also the reason that the admittance matrix is always very sparse. Since each bus in a real power system usually only connects to a few other buses through transmission lines, even if the system's scale is very large. Furthermore, each diagonal element approximately equals the negative of the sum of off-diagonal elements in the corresponding row/column, and the difference is caused by the shunt admittance and the tap-ratios of transformer lines. In other words, the sum of each row/column is close to zero, or exactly zero if these is no shunt admittance at the corresponding bus and no transformer connected to the bus. In the mathematical field of graph theory, the Laplacian matrix, also called admittance matrix, is a matrix representation of a graph. It is equal to the graph's degree matrix minus the adjacency matrix. For an undirected graph, which is applicable to power systems, the Laplacian matrix is symmetrical, and each row's or each column's elements summation is zero. So, the power system and the undirected graph are closely mapping to each other, indicating the feasible applications of GDB into power systems.

### *C. Graph Computing and Its Application in Power Flow Analysis*

*1) Node-Based Parallel Computing*: In graph computing, each node is independent to others and capable of conducting the local computation. Using the mode of all node synchronization by activating all nodes at the same time, the node-based graph operation is implemented in parallel to save computation cost and improve the computation efficiency. This paper proposes to apply graph computing technology into power flow analysis. Take the admittance matrix as an example, off-diagonal elements are locally and independently calculated based on the admittance attributes of the corresponding edges, and each diagonal element is obtained by processing the admittance attributes in the node and its connected edges. So, the admittance matrix can be computed in parallel within one graph traversal. Other examples of node-based parallel computation in power flow analysis are each power flow iteration's active and reactive power injection calculation, node-based variables mismatch comparison and convergence check, and the post-convergence active and reactive branch power flow calculation.

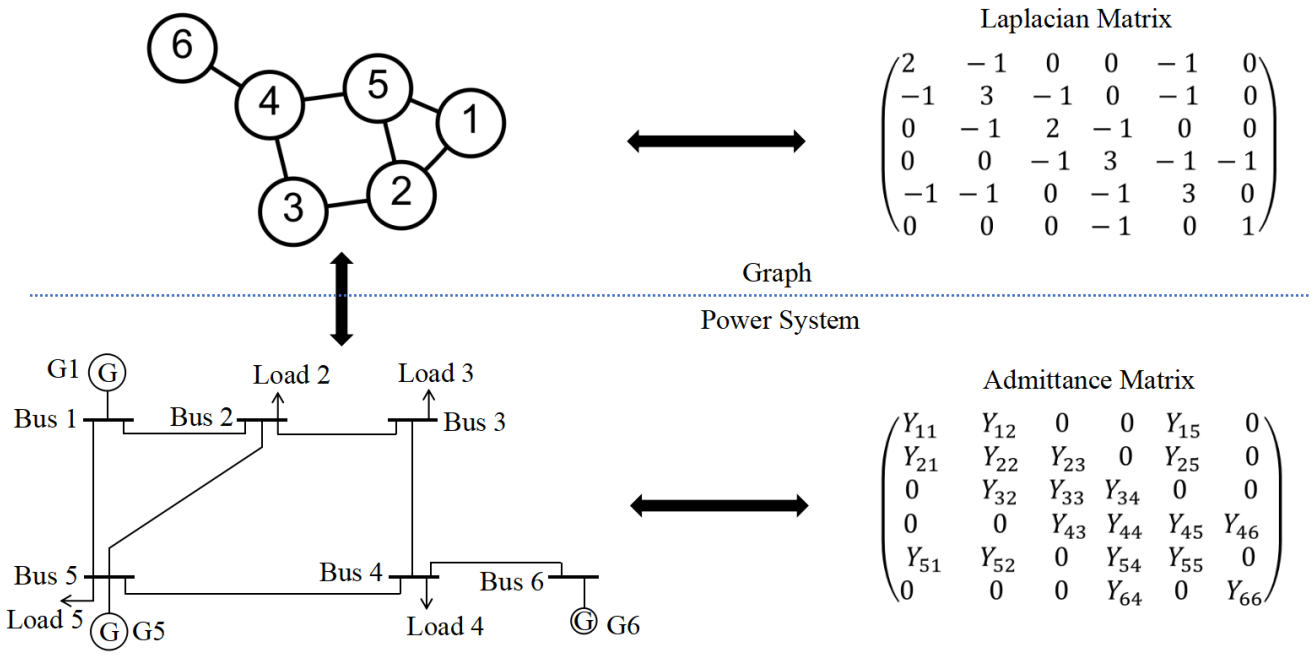

Figure 2. Mapping between graph and power system

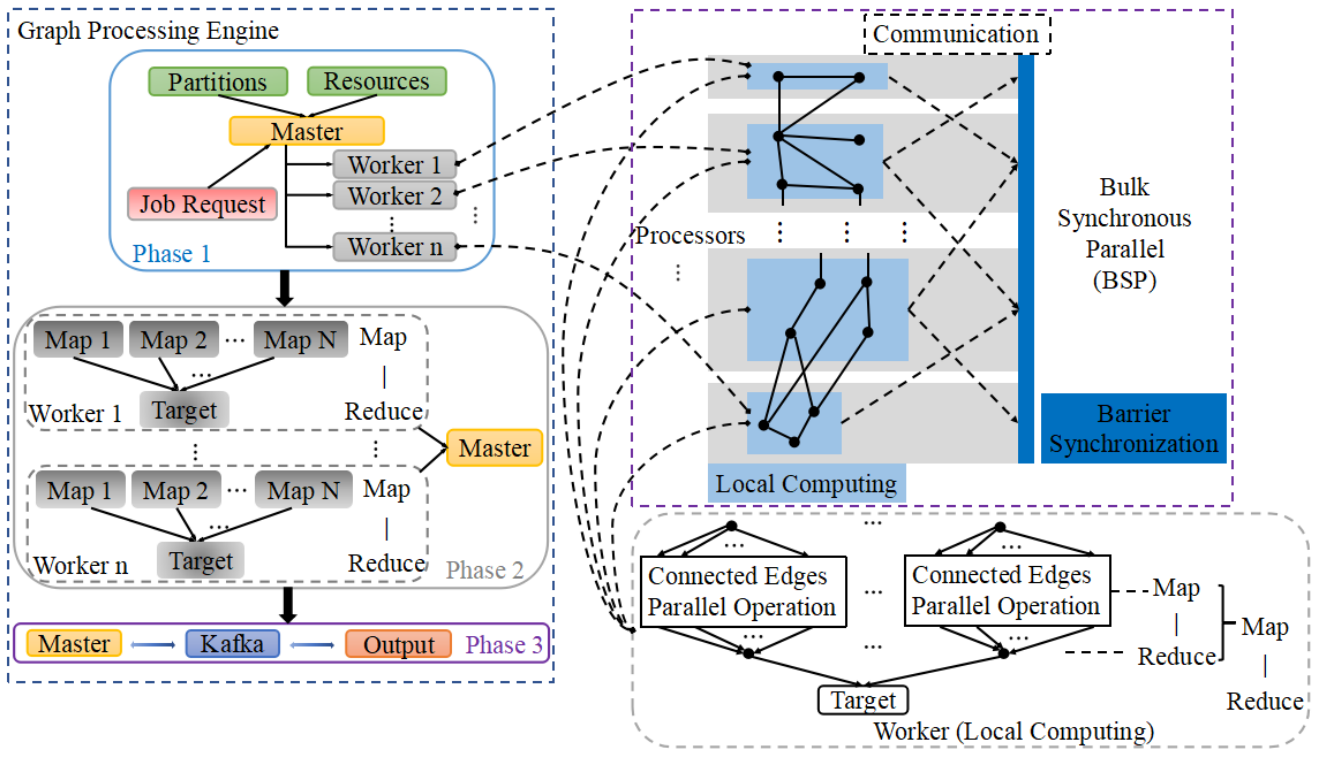

Figure 3. MapReduce and BSP inside graph processing engine

*2) MapReduce and Bulk Synchronous Parallel:* In computing, MapReduce and bulk synchronous parallel (BSP) are two major parallel computation models. BSP is a bridging model for designing parallel algorithms. BSP consists of components who are capable of local memory transactions, a network that communicates messages between components, and a facility allows for synchronization of components. As presented in Fig. 3, within the graph processing engine (GPE), the master processor assigns tasks to worker processors per the CPU resources, data partitions and job request. Each worker focuses on its local computation, communicates with other workers, and outputs results during barrier synchronization. This process is implemented in BSP. For each worker, it employs MapReduce scheme to do local logic and algebraic operation in parallel. MapReduce is a framework of processing massive datasets in form of <key; value> pairs and plays a prominent role in parallel computing. It includes two phases, map phase, performing local data processing in parallel, and reduce phase, processing output data per key in parallel. Below is the MapReduce programming mechanism in graph computing. Using SELECT syntax, MapReduce processes are generated for selected nodes. Each node's MapReduce is procressed in parallel. Beginning from ACCUM syntax, map phase starts to do edges operations for the corresponding node. In the POST-ACCUM, reduce phase updates and aggreates results for the node.

**MapReduce in Graph Computing**

```
Initialize T0 = {all nodes};
T1 = SELECT s FROM T0:s-(edges:e)−>t
  // Start MapReduce processes for selected nodes
ACCUM
[edge operations]
  // map phase for each selected node
POST-ACCUM
[vertex operations];
  // reduce phase for each selected node
End;
```

## III. Bi-Level PageRank Algorithm for Power Flow Analysis Using Graph Database

### *A. Graph Computing based PageRank*

PageRank is an algorithm, used by Google Search, to rank websites by calculating the importance of the web pages. Its equation is as follows, indicating the PageRank of page $p_i$ at time point of $(t+1)$.

$$PR(p_i;t+1)=\frac{1-d}{N}+d\cdot\left(\sum_{p_j\in B_{p_i}}\frac{PR(p_j;t)}{L(p_j;t)}\right) \quad (1)$$

where $d$ is the damping factor, generally around 0.85, $B_{p_i}$ is the set of pages linked to $p_i$, $L(p_j;t)$ is the number of out links on page $p_i$, and $N$ is the number of pages.

Equation (1) presents two characteristics of PageRank algorithm: (a) each page's PageRank value only depends on its neighboring pages' PageRank values and the number of its out links; (b) during each iteration, the calculation of the PageRank for each page only uses the values obtained from a previous iteration. The former discloses that the PageRank of each page can be locally calculated, and the latter character implies that PageRank algorithm can be implemented in parallel. Therefore, based on the description of graph computing in the last section, in each iteration of PageRank, node-based parallel computing is applicable.

### *B. Bi-Level PageRank Algorithm in Power Flow Analysis with Graph Computing*

In power systems, the power flow equations are presented in (2). Using Jacobi method, the power flow iteration is shown in (3), which is very similar to the PageRank algorithm. The node voltage depends on its neighboring nodes' voltages, the line admittance between itself and neighboring nodes, and its own attributes, like voltage phasor, node admittance, active power injection and reactive power injection. Furthermore, its calculation is determined by values achieved from last iteration. Therefore, the power flow analysis is fully feasible in parallel computing. But the slow rate of convergence is still an issue, even though it takes little memory and does not need to solve matrix. To improve its convergence, this paper employs two strategies: (a) using damping factor; (b) separating nodes into two levels.

With the addition of the damping factor, equation (3) is developed into (4). Like the function in PageRank algorithm, damping factor is used to improve the convergence of power flow calculation. First, it could avoid a sink when zero-impedance branch exists. Look at (3), if there is a zero-impedance branch connected to node $i$, the values of $\sum_{\substack{j=1 \\ j\neq i}}^{n} Y_{ij}\dot{V}_j$ and $Y_{ii}$ are too large to reflect power changes in voltage update, leading to a voltage sink at node $i$, and a worse convergence. In addition, with the help of damping factor, the frequency of back and forth fluctuations around the real system state could be much reduced to improve the convergence. But, the convergence is still very slow with the use of damping factor, especially when a very high precision is pursued. This is determined by the algorithm itself, in which each iteration's calculations only depend on results obtained in the previous iteration. Borrowing the core idea of Gauss-Seidel to improve power flow convergence, this paper proposes a bi-level PageRank algorithm to improve convergence and meanwhile maintain the parallel computing. The system nodes are divided into two levels in the GDB, ensuring, in each level, most of nodes are not mutually connected. So, based on (4), equation (5) is developed for the bi-level PageRank based power flow analysis. In this way, the corresponding graph for this power system is also divided into two levels, $A$ and $B$. With the help of node-based graph computing, level $A$ is first updated using results from a previous iteration. Then nodes in level $B$ are updated using level $A$'s results in this iteration and level $B$'s information from last iteration. The computation procedure in graph computing is also displayed below.

$$\dot{S}_i^{\,*} = P_i - jQ_i = \dot{V}_i^{\,*}\sum_{j=1}^{n}(Y_{ij}\cdot\dot{V}_j)$$

$$\rightarrow\ \dot{V}_i = \frac{1}{Y_{ii}}\left[\frac{P_i - jQ_i}{\dot{V}_i^{\,*}} - \sum_{\substack{j=1 \\ j\neq i}}^{n} Y_{ij}\dot{V}_j\right] \tag{2}$$

$$\dot{V}_i^{(k+1)} = \frac{1}{Y_{ii}}\left[\frac{P_i^{(k)} - jQ_i^{(k)}}{\dot{V}_i^{(k)*}} - \sum_{\substack{j=1 \\ j\neq i}}^{n} Y_{ij}\dot{V}_j^{(k)}\right] \tag{3}$$

$$\dot{V}_i^{(k+1)} = (1-d)\dot{V}_i^{(k)} + \frac{d}{Y_{ii}}\left[\frac{P_i^{(k)} - jQ_i^{(k)}}{\dot{V}_i^{(k)*}} - \sum_{\substack{j=1 \\ j\neq i}}^{n} Y_{ij}\dot{V}_j^{(k)}\right] \tag{4}$$

$$\dot{V}_i^{(k+1)} = \tag{5}$$

$$\begin{cases} (1-d)\dot{V}_i^{(k)} + \frac{d}{Y_{ii}}\left[\frac{P_i^{(k)} - jQ_i^{(k)}}{\dot{V}_i^{(k)*}} - \sum_{\substack{j\in A \\ j\neq i}} Y_{ij}\dot{V}_j^{(k)} - \sum_{j\in B} Y_{ij}\dot{V}_j^{(k)}\right] \\ \qquad (if\ \ i\in A) \\ (1-d)\dot{V}_i^{(k)} + \frac{d}{Y_{ii}}\left[\frac{P_i^{(k)} - jQ_i^{(k)}}{\dot{V}_i^{(k)*}} - \sum_{j\in A} Y_{ij}\dot{V}_j^{(k+1)} - \sum_{\substack{j\in B \\ j\neq i}} Y_{ij}\dot{V}_j^{(k)}\right] \\ \qquad (if\ \ i\in B) \end{cases}$$

**Graph Computing based Bi-Level PageRank Algorithm in Power Flow Analysis**

```
Initialize T0 = {all nodes}
T1 = SELECT s FROM T0:s-(edge:e)−>t
     ACCUM
     [calculate off-diagonal elements in Ybus matrix],
     [sum up off-diagonal elements for each node in Ybus matrix]
     POST-ACCUM
     [complete diagonal elements calculation for Ybus matrix],
     [initialize system states];
while (Re{V} > threshold & Im{V}>threshold){
T2 = SELECT s FROM T1:s-(edge:e)−>t
     ACCUM
     [calculate Y_ij V_j^(k) through edge operations]
     POST-ACCUM
     [update voltages via node operations],
     [update power mismatch and voltage changes from last iteration];}
End;
```

## IV. Case Study and Discussion

### A. Testing Environment and Testing Cases

In this paper, the cases are tested on a server with the graph computing platform, which is TigerGraph v0.8.1. Detailed testing environment is presented in Table I. Besides, the testing cases include IEEE 14-bus system, IEEE 118-bus system, MP 10790 system, and the extension cases, MP 10790*10 and MP 10790*100. Furthermore, a real provincial system with 1425 buses and 1922 branches, FJ case, is employed to further verify the proposed algorithm's effectiveness and high performance.

### B. Algorithm Verification

To verify the graph computing based bi-level PageRank algorithm's accuracy in power flow analysis, Table II provides the results comparison between the proposed algorithm and MatPower, using IEEE 118-bus system. Regarding the proposed algorithm, the convergence criteria depends on the maximum real part and imaginary part of voltage phasors. In MatPower, its convergence criteria are based on the maximum

TABLE I. Test Environment

| Hardware Environment | |
|---|---|
| CPU | 2 CPUs × 6 Cores × 2 Threads @ 2.10 GHz |
| Memory | 64 GB |
| **Software Environment** | |
| Operation System | CentOS 6.8 |
| Graph Database | TigerGraph v0.8.1 |

TABLE II. IEEE 118-BUS SYSTEM RESULTS COMPARISON

| Method | | Graph Computing based Bi-Level PageRank | | Matpower |
|---|---|---|---|---|
| | | Precision of 0.00003 | Precision of 0.00025 | Precision of 0.05 |
| Size of LCB | | 128 | 128 | — |
| # of LCB | | 1 | 1 | — |
| # of Running Threads | | 1 | 1 | 1 |
| Iterations | | 342 | 383 | 329 |
| Computation Time (ms) | | 174.25 | 227.91 | 820 |
| Max Difference from MatPower | Angle (radian) | 0.0245 | 0.0133 | — |
| | Magnitude (per unit) | 0.0006 | 0.0005 | — |

mismatches of real power and reactive power at each node, and the precision is set at 0.05.

In Table II, it can be clearly seen that, from the view of the convergence and calculation performance, the number of iterations is in the same level as MatPower, which is determined by the algorithm itself. But the calculation time is less than 30% of the time spent in MatPower. Each iteration only costs about 0.5~0.6 ms. Regarding the results difference, the maximum angle difference is around 1.40 degree, which is about 0.0244 in radian, with the converge criterion of 0.0003. If the converge criterion is set as 0.00025, the maximum angle difference is around 0.76 degree, which is about 0.0133 in radian. The maximum magnitude difference is very small and can be neglected under both convergence criterion. In addition, the computation time in Table II presents that the proposed algorithm has much better performance than Matpower, with one running thread.

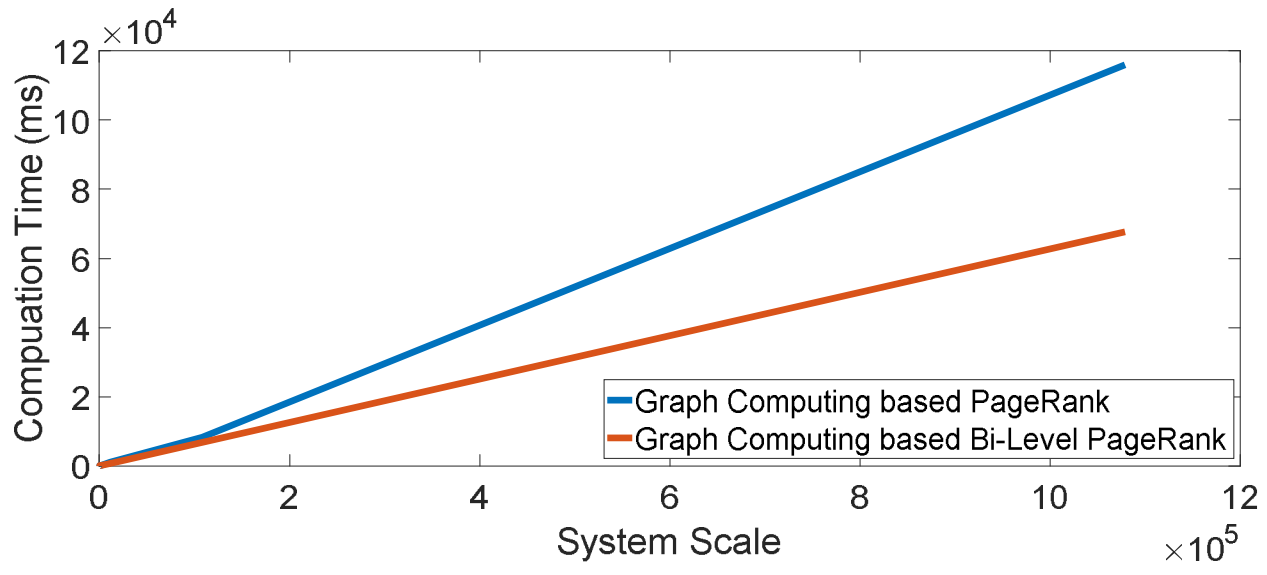

Figure 4. Computation time using graph computing based PageRank and bi-level PageRank in different system scales

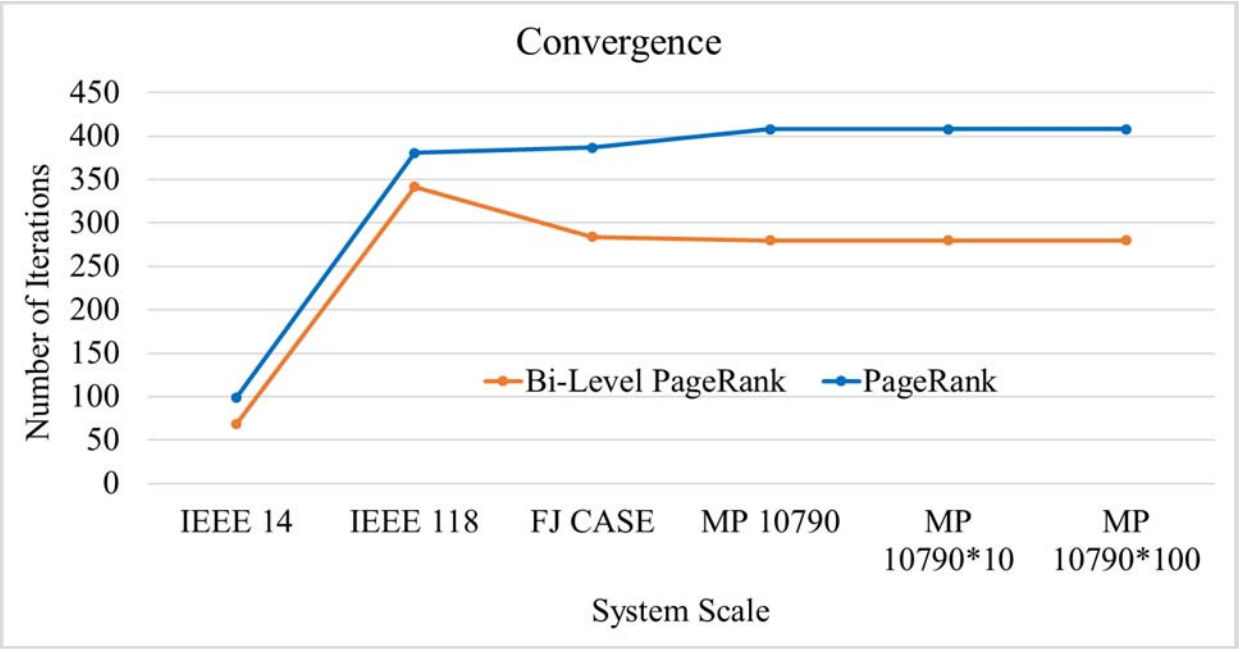

Figure 5. Convergence comparison of PageRank and bi-level PageRank algorithms in different systems

### C. *Comparison between PageRank and Bi-Level PageRank*

In this section, the performance comparison between PageRank and the proposed bi-level PageRank is presented in Fig. 4 and Fig. 5.

Based on the results, in the small-scale case, the computation time using the basic PageRank method is close to, even better than, the bi-level PageRank algorithm. This is because, in the small-scale case, the graph traverse costs little and there is not much difference in the convergence performance between the two algorithms, while the bi-level PageRank algorithm spent extra time on communication and data update between levels. However, in the large-scale case, the time spent on graph traverse costs much more and dominates the computation time. With the proposed bi-level PageRank algorithm, the number of iterations significantly decreases, and the convergence is improved ~30%.

### D. *Parallelism Testing*

In this section, cases of FJ system, MP 10790, MP 10790*10, and MP 10790*100 are tested with different running threads to demonstrate the high-performance parallelism based on graph database.

As shown in Table III, FJ case has the best performance with 12 running threads, while the rest three cases have least computation time with 16 running threads. This is because FJ case system-scale is much smaller than the other three cases. In addition, since the server only has 12 cores, as shown in Table I, the computation time of MP 10790, MP 10790*10, and MP 10790*100 decreases slowly when the number of running threads gets close to 12 and hits the best result at 16 running threads. In Fig. 6, the curves of computation time vs. running threads indicate that, with the use of multi-thread in GDB, the computation speeds up, and as the system scale increases, the parallelism performance is more obvious.

## V. CONCLUSION AND FUTURE WORK

This paper applied graph database to power system modeling and proposed a GDB based bi-level PageRank algorithm for power flow analysis by making use of node-based parallel computing. Cases, including a real provincial system in China, MP 10790 and its extension systems, are studied to verify the proposed approach's accuracy and high performance computation efficiency.

TABLE III. GRAPH COMPUTING BASED BI-LEVEL PAGERANK PARALLELISM TESTING

| Threads \ System | FJ system | MP 10790 | MP 10790*10 | MP 10790*100 |
|---|---|---|---|---|
| 1 | 654.71 | 4308.3 | 47352 | 493611 |
| 2 | 483.29 | 2673.8 | 28231 | 281074 |
| 4 | 357.40 | 1561.2 | 14635 | 145611 |
| 8 | 294.23 | 991.62 | 8356.8 | 83983 |
| 12 | **287.90** | 795.66 | 6978.4 | 72410 |
| 16 | 313.10 | **766.44** | **6841.4** | **67753** |
| 20 | 316.67 | 893.03 | 6930.6 | 69511 |
| Speed-Up | 2.27 | 5.62 | 6.92 | 7.29 |

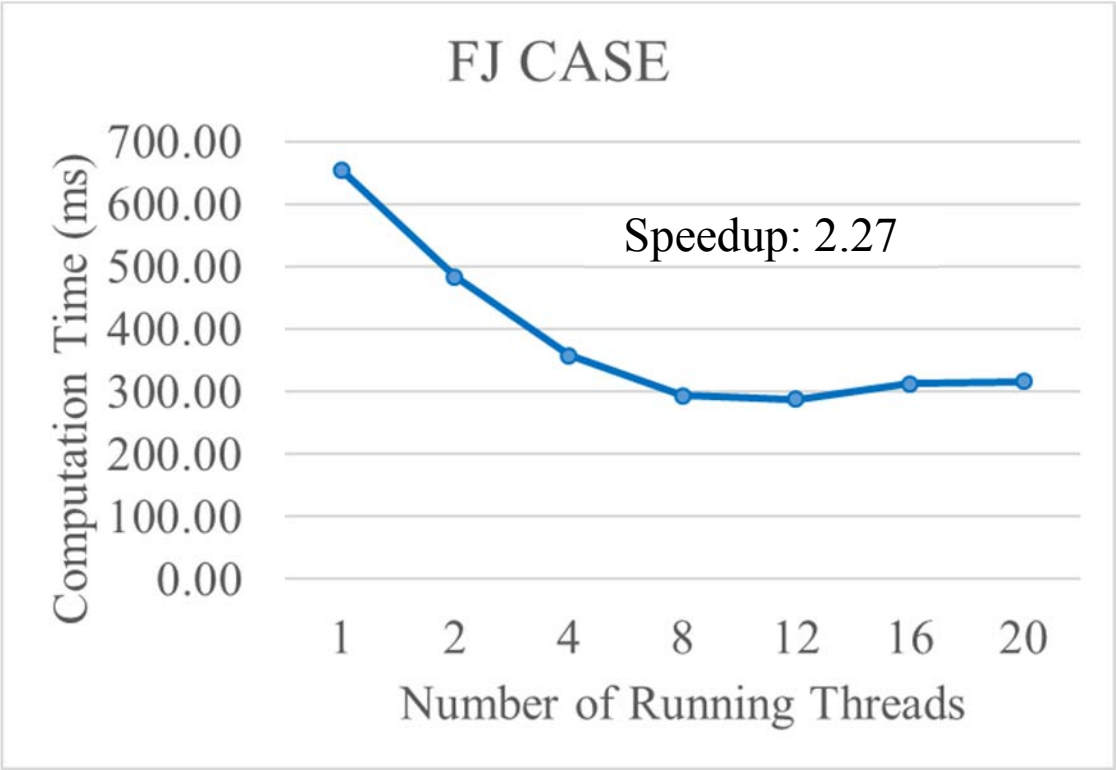

(a)

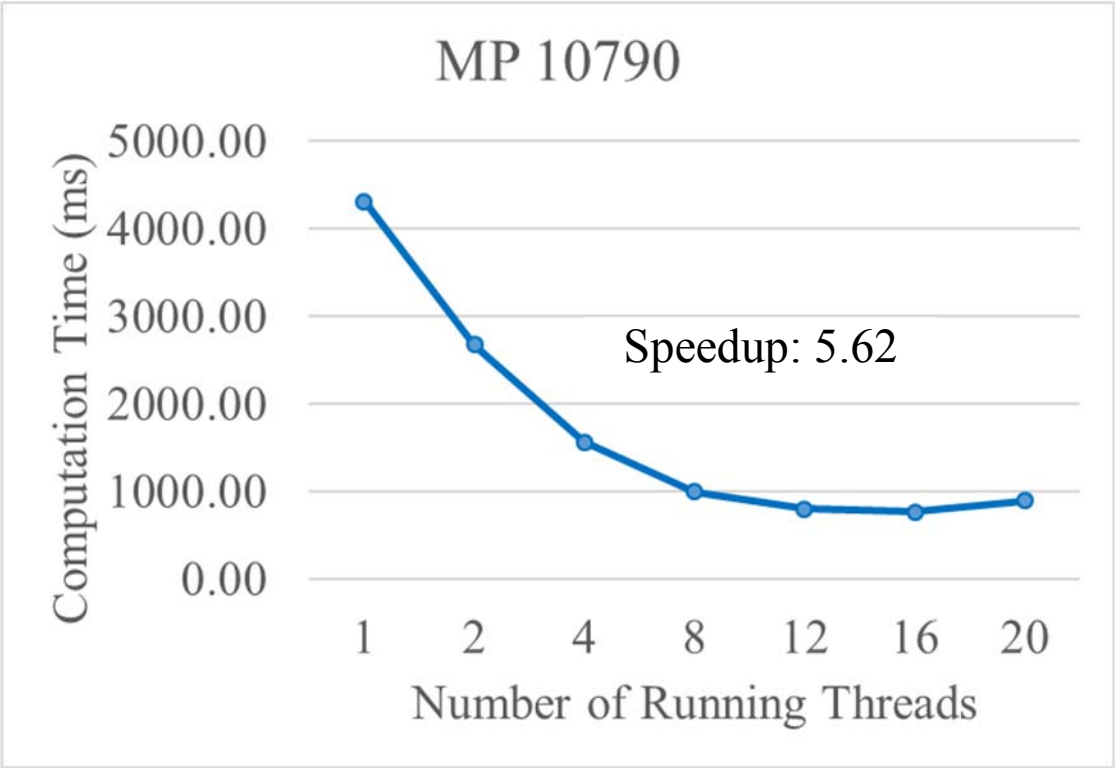

(b)

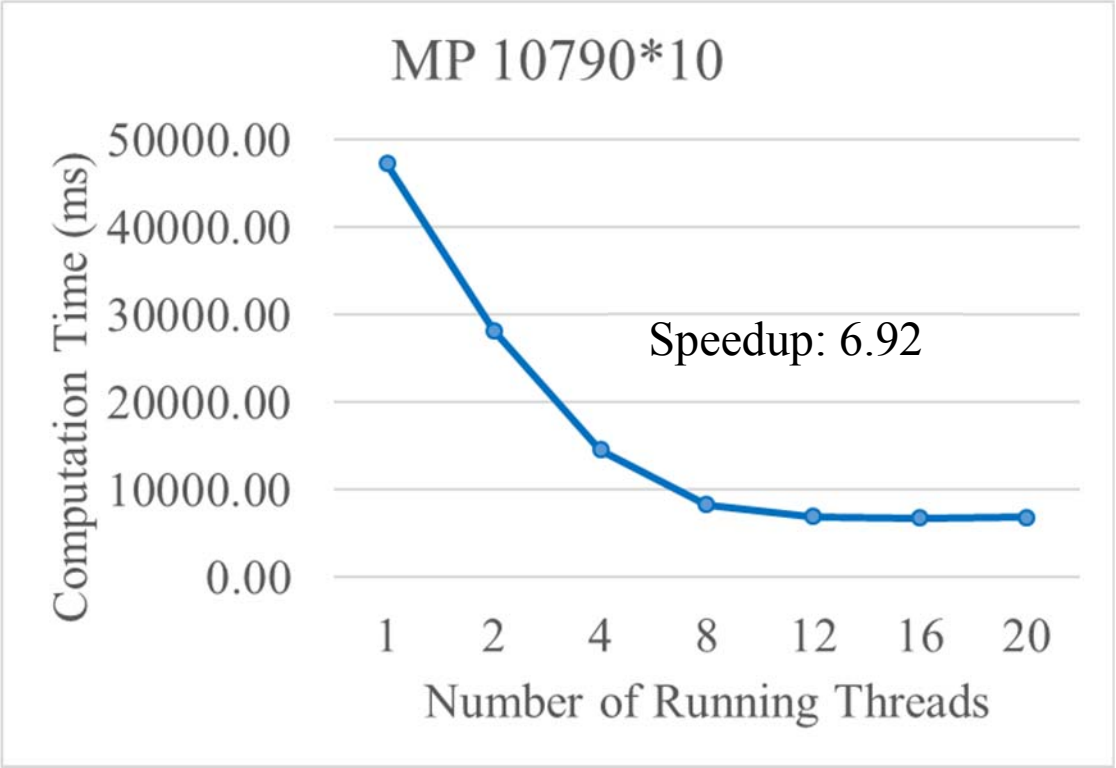

(c)

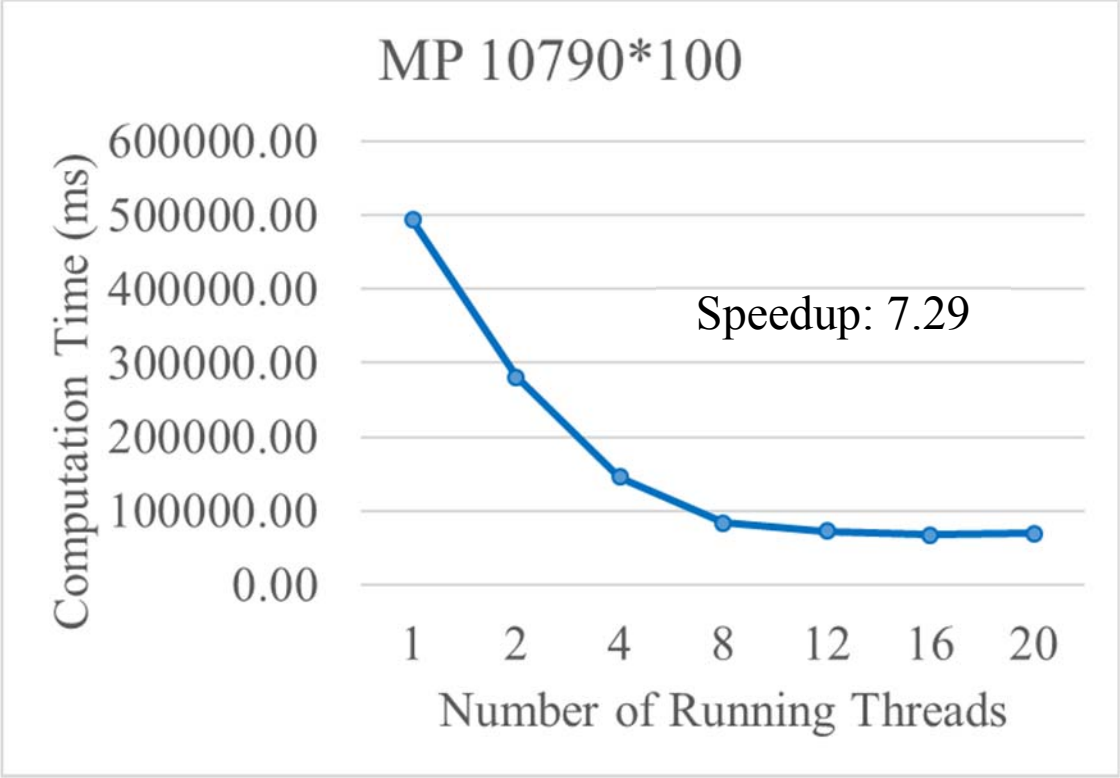

(d)

Figure 6. Parallelism testing in graph database